\documentclass{aa}
\usepackage[varg]{txfonts}
\usepackage{graphicx}
\usepackage{amsmath,amssymb,pgf,graphicx,color,url}
\usepackage[colorlinks=true,citecolor=blue,linkcolor=blue]{hyperref}

\begin{document}
\title{Search for glitches of gamma-ray pulsars with deep learning}
\author{E.V.~Sokolova\inst{\ref{inst1}}\thanks{E-mail: \email{sokol@ms2.inr.ac.ru}}
\and A.G.~Panin\inst{\ref{inst1},\ref{inst2}}\thanks{E-mail: \email{panin@ms2.inr.ac.ru}}}
\institute{Institute for Nuclear Research of the Russian Academy of Sciences,
  Moscow 117312, Russia\label{inst1}
  \and Moscow Institute of Physics and Technology, Dolgoprudny 141700,
  Russia\label{inst2}}

\date{Received <date> /
  Accepted <date>}

\abstract {The pulsar glitches are generally assumed to be an apparent
  manifestation of the superfluid interior of the neutron stars. Most of
  them were discovered and extensively studied by continuous monitoring
  in the radio wavelengths. The {\it Fermi}-LAT space telescope has made a
  revolution uncovering a large population of gamma-ray pulsars. In this
  paper we suggest to employ these observations for the searches of new
  glitches. We develop the method capable of detecting step-like frequency
  change associated with glitches in a sparse gamma-ray data. It is based
  on the calculations of the weighted $H$-test statistics and glitch
  identification by a convolutional neural network. The method demonstrates
  high accuracy on the Monte Carlo set and will be applied for searches of
  the pulsar glitches in the real gamma-ray data in the future works.}

\maketitle

\section{Introduction}
Pulsars are fast-rotating highly magnetized neutron stars.
Rotating at a frequency
gradually decreasing over long time due to radiation
they rightfully deserve the status of the most precise clocks in the Universe.
However, its stability is violated by glitches. Pulsar glitch manifests
itself as sudden step-like increase of rotation frequency
and its time derivative. Glitch can be characterized by the
so-called size~---~the relative frequency change. Detected glitch sizes
vary from very small values $\Delta f/ f \sim 10^{-12}$~\citep{McKee}
comparable with timing noise to the largest
$\Delta f/ f \sim 10^{-5}$~\citep{Espinoza}. For example, the Vela pulsar
experiences quite large glitches of the size $10^{-6}$ approximately every
1000 days~\citep{Cordes, Shannon, Packard:2018}, while small frequency
changes of the size less than $2\times 10^{-7}$ were demonstrated by the
Crab pulsar~\citep{Espinoza:2014, Lyne}.

Although the first glitch was discovered more than fifty years
ago~\citep{Manchester, Downs} (see, e.g.~\citet{Vivekanand} for a review),
the exact origin of these phenomena is still open
to debate~\citep{Haskell:2015jra}. Initially, glitches were
associated with starquakes~\citep{Ruderman}, but
then the superfluid model was put forward to explain this
phenomena~\citep{Packard:1972}. With the discoveries of new glitches we
become closer to understanding their nature, which in turn may shed light on
the internal structure of the neutron stars~\citep{Espinoza:2014}.

The radio surveys produced most of the glitch discoveries due to the longest
accumulated observation times and the largest number of observed pulsars (see
the ATNF pulsar
catalog\footnote{https://www.atnf.csiro.au/research/pulsar/psrcat/} and the
JBO online glitch
catalog\footnote{http://www.jb.man.ac.uk/pulsar/glitches.html}).
However, some of the pulsars are radio-quiet, observable only in gamma-ray band
with no radio counterpart. Before the launch of the
{\it Fermi Gamma-ray Space Telescope} with Large Area Telescope (LAT) on
board in 2008~\citep{Atwood},
there was known only one such object~---~Geminga~\citep{Halpern, Bertsch}.
Now more than 250 LAT sources are identified as gamma-ray
  pulsars\footnote{http://tinyurl.com/fermipulsars}~\citep{TheFermi-LAT:2013ssa}
  and more than 50 among them are radio quiet. Observations in gamma-rays
  may provide reach information about glitches and have the potential to
  search for the difference in their properties of radio-quiet and radio-loud
  populations.

Large fraction of the LAT-detected gamma-ray pulsars are young and
  energetic. Recent studies suggest that young pulsars experience glitches
  more often then the old ones~\citep{McKenna, Lyne:2000, Espinoza}.
  Several glitches of the size of the order of $10^{-5}$ are already
  discovered simultaneously with the discovery of the pulsars itself via blind
  searches~\citep{Abdo:2009, Saz:2010, Pletsch:2012, Pletsch:2013, Clark:2015}.
  This gives us hope to identify more new glitches in a targeted extensive
  search in the {\it Fermi}-LAT data.

Detection of glitches in the sparse gamma-ray data is computationally
challenging. The lack of rigorous criteria to distinguish glitches from other
peculiarities at low signal-to-noise ratio compels to search them manually.
In this paper we suggest a method which helps to identify glitches
automatically. It is based on the computations of the weighted $H$-test
statistic~\citep{Jager:1989,deJager:2010ci} widely used in the blind searches
of new gamma-ray pulsars and the glitches analyses~\citep{Clark:2017}.
In order to recognize glitches in the resulting data we suggest to apply
the machine learning techniques. It is a modern tool which has already
found a lot of applications in a broad range of astrophysical
problems~\citep{BallBrunner,Baron} including selection of radio pulsar
candidates~\citep{Eatough}. In this paper using the Monte Carlo
data we have show that a convolutional neural network is capable to find
pulsar glitches of different sizes with the high accuracy. We plan an
extensive applications of the method to the real data in the future works.

\section{Method}
\label{sec:method}
In this section we describe the method used to detect glitches of the 
gamma-ray pulsars. We apply it to the {\it Fermi} Large Area Telescope
  data prepared with {\it Fermi Science Tools} package according
  to~\citet{Sokolova:2016btd}.
The data consist of individual photons recorded between 2008 August 4
(MJD 54682) and 2015 March 3 (MJD 57084) selected from "SOURCE" class events
of the Reprocessed Pass 7 data set by the {\it gtselect} tool according to the
following criteria. Photons were included if they had energy above $100$
MeV, arrived within $8^\circ$ of a target source, with a zenith angle
$< 100^\circ$ and when the LAT’s rocking angle was $< 52^\circ$.

A source model which includes {\it Fremi}-LAT 3FGL sources in a $8^\circ$
radius circle, galactic and isotropic diffuse emission components is
constructed for each of the pulsar considered in the paper. The model
parameters were optimized with unbinned likelihood analysis by the
{\it gtlike} tool. Next, using the {\it gtsrcprob} tool each photon is
assigned a weight --- probability to be emitted by the source. To search
for glitch $40000$ photons with the highest weights were kept for each pulsar.

  The glitches search method employs the photon arrival times $t_i$
  at the solar system barycenter. Position-dependent “barycentering”
  corrections were calculated by the {\it gtbary} tool. These corrections
  take into account the Earth's orbital motion which causes Doppler
  modulation of pulsations and complicates the search for glitches.

At the first step, we combine the data into several time
groups which contain photons within the $170$-day time window sliding over
the entire data with the $17$-day time step. Working with $6.5$ years of
observations we prepared in this way $131$ time groups of the photons.
The particular choice of a time-window size and a sliding step were
suggested by~\citet{Pletsch:2013} as a balance between the signal-to-noise
ratio and the time resolution of the method. Then the photon arrival times are
corrected to compensate for the frequency evolution,
\begin{equation}
  \label{time}
\tilde{t}_i = t_i+\frac{\gamma}{2}(t_i-t_0)^2\;,
\end{equation}
where $\gamma = \dot{f}/f$ and $t_0 = 286416002$ (MJD 55225) is a reference
epoch. Then, the value of $H$ is computed separately for each group of
the photons according to the formula
\begin{equation}
\label{H}
H = \max_{1 \leq L \le20}\left[\sum_{l=1}^{L}\mid\alpha_{l}\mid^{2}-4(L-1)\right]\;,
\end{equation}
where $\alpha_l$ is a Fourier amplitude of the $l$-th harmonic,  
\begin{align*}
  \alpha_{l} &= \frac{1}{\varkappa}\sum_{i}w_{i}\mathrm{e}^{-2\pi i l f \tilde  t_i}\;,\\
  \varkappa^2 &= \frac{1}{2}\sum_{i}w_{i}^2\;.
\end{align*}
Fourier exponents in this formula are multiplied by the
photon weights $w_i$.

\begin{figure}
  \resizebox{\hsize}{!}{\includegraphics{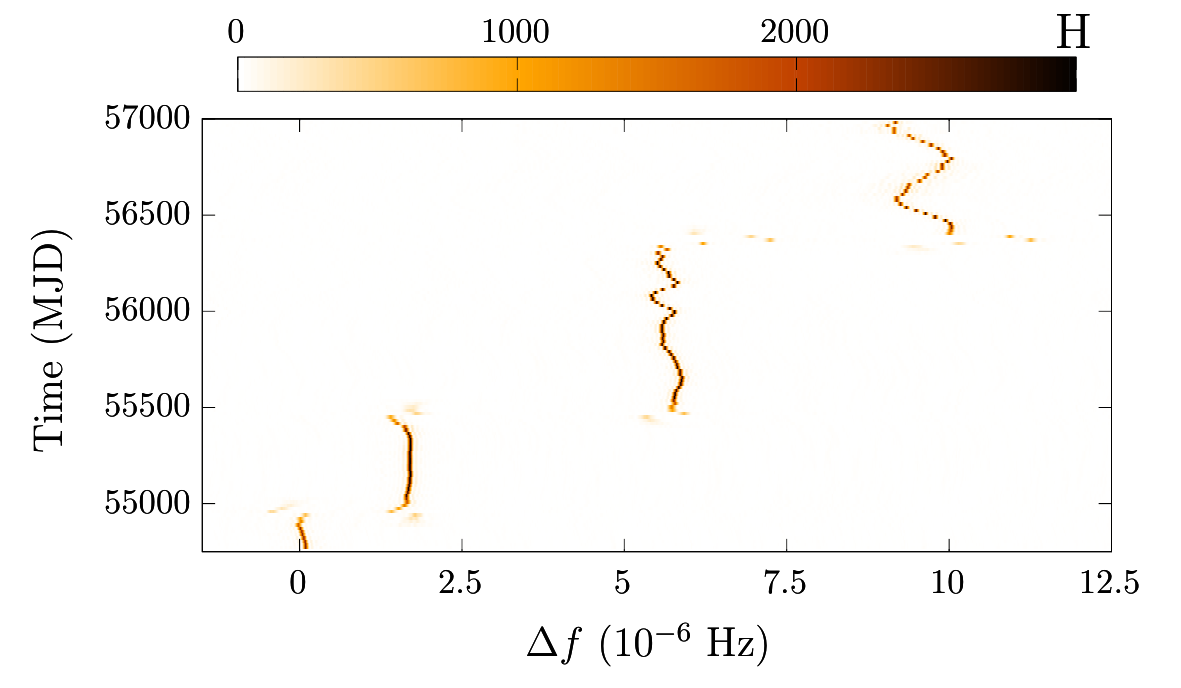}} 
  \caption{Pulsar glitch analyses for PSR J0007+7303. The weighted $H$-test
    is calculated according to the equation ~\eqref{H} using photons within the
    170-day time window slided over the entire data set with the 17-day step.
    For each window scans over $f$ and $\dot{f}$ are done. The maximal value
    of $H$ over $\dot{f}$ for a given $f$ is shown by color. Vertical axis
    shows the time midpoint of each time window. Horizontal axis shows the
    offset in $f$.}
   \label{fig1}
\end{figure}

Weighted $H$-test staticstics is a powerfull tool to search for weak
  periodic signals with unknown light curve shape in sparce data.
  It tests whether photon phases calculated by folding the arrival times
  $t_i$ at a given frequency (and at a given spin-down rate, which enters
  into equation~\eqref{time}) are uniformly distributed. Otherwise a periodic
  signal with spin-parameters $f$ and $\dot{f}$ presents in the data and its
  significance is given by $H$. Consequently, the correct values of frequency
and spin-down rate if unknown a priori can be determined
as a maximum of the $H$-test by scanning over it in some range. Performing it
separately for each of the time groups of the photons introduced above we
obtain data describing the dependence of $H$-test on time, frequency and
spin-down rate. In order to reduce the size of the data sample we
maximize the $H$-test over $\dot{f}$ at a fixed frequency and time and
finally obtain the dependence $H(f,t)$. Exploring these results one can
detect an abrupt change of frequency associated with a glitch.   

To demonstrate the method we apply it to PSR J0007+7303 in the {\it Fermi}-LAT
data. The result is presented in Figure~\ref{fig1}. The color code represents
the weighted $H$-test maximized over the spin-down rate $\dot{f}$. Vertical
axis shows the time midpoint of each time group of the photons introduced
above. One may see from the figure that frequency position of the maximum
$H$ changes abruptly over time revealing three pulsar glitches around 55000
MJD, 55500 MJD and 56400 MJD~\citep[for more details see][]{Li:2016uxb}.

As one can see from Figure~\ref{fig1}, the maximum of the weighted
  $H$-test between glitches oscillates around a certain frequency. This is
  due to inaccurate 3FGL sky-coordinates of the source limited by the LAT’s
  angular resolution to a few arc minutes. As a result, incorrectly
  taken into account satellite's motion relative to the source leads to
  the Doppler modulation of pulsations.

Figure~\ref{fig1} gives an example of the large Vela-type glitches of the size
$\Delta f/ f \sim 10^{-6}$ clearly visible by the naked eye in the $H$-test
data despite the Doppler frequency shift. However, identification
of small glitches of the size $\Delta f/ f \sim 10^{-8} - 10^{-7}$ for the
gamma-ray pulsar may become a difficult problem especially if the
pulsars are not so bright.

\begin{figure}
  \resizebox{\hsize}{!}{\includegraphics{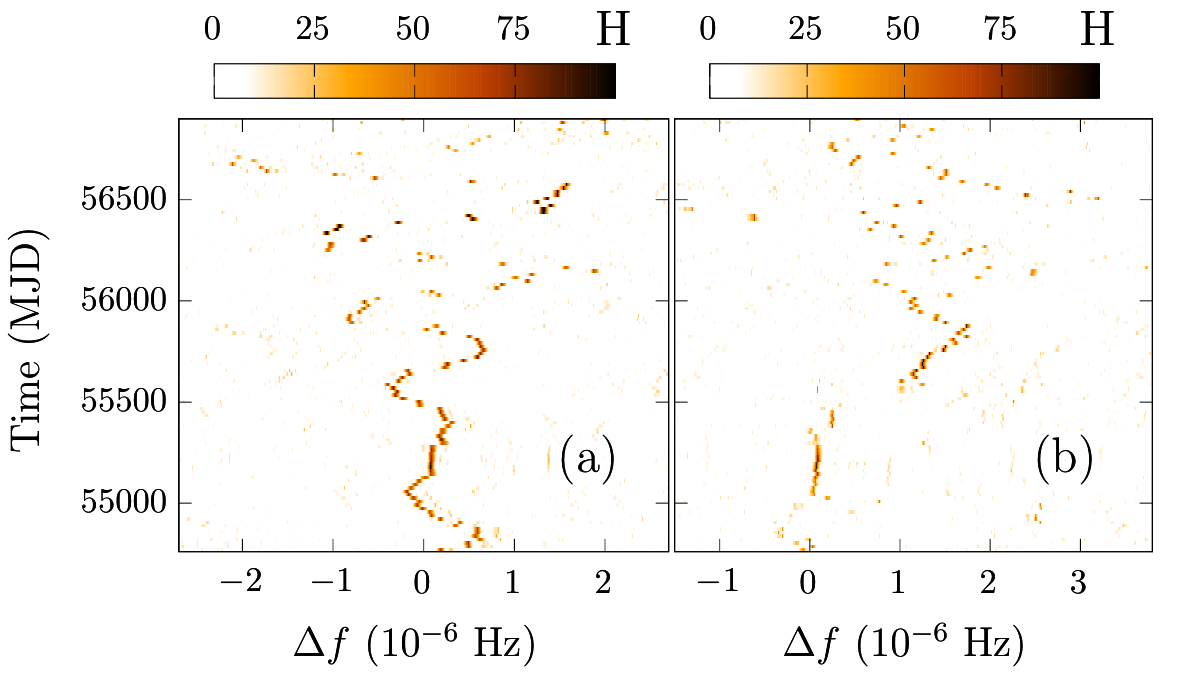}}
  \caption{The same as in Figure~\ref{fig1} for: (a) PSR J2030+3641
    without glitches during considered epochs and (b) PSR J1422-6138
    which experienced glitches at 55310 MJD and 55450 MJD. Computations were
    performed using coordinates of the sources from the {\it Fermi}-LAT 3FGL
    catalog.}
  \label{fig2}
\end{figure}

Another source of coherence loss is the poor background rejection
  when the pulsar is not very bright so that a large number of selected
  photons were not actually emitted by the pulsar.
Figure~\ref{fig2} gives an example of such pulsars: PSR J2030+3641
(see Figure~\ref{fig2}a) without glitches during the considered period
and PSR J1422-6138 (see Figure~\ref{fig2}b) which experienced two
glitches~\citep{Pletsch:2013}. In the computations we used coordinates
from the {\it Fermi}-LAT 3FGL catalog, therefore the Doppler shift
of the frequency presents in the $H$-test data. All that makes glitches of the
pulsar shown in Figure~\ref{fig2}b hardly distinguishable in the background
of other frequency distortions.

More accurate coordinates of the pulsar can be used in the analysis,
  if they are known from other observations. If not, the loss of
  phase-coherence can be reduced by refining the sky-location of the source
  \citep{Yu:2013, Pletsch:2013}.
However, it enlarges the parameter space of the scan to four-dimensional
(sky position, frequency and spin-down rate).
A more efficient selection of the photons have to be performed
  in the case of large background radiation causing frequency distortions
  in the $H$-test data. As a result, many samples corresponded to each
  attempt of the scan over coordinates and/or photons selection will be
  available for the analysis. Visual inspection all of them is rather
  difficult. Therefore a reliable criterion for automatic glitches search
  which are able to "look at" each sample is needed.

\section{Convolutional neural network}
\label{sec:neural}
Efficient glitches identification from other peculiarities in the $H$-test
data can be obtained by using the machine learning approach. It provides the
ability to “learn” specific patterns corresponding to the pulsar glitch directly
from the data, without being explicitly programmed. In the present paper we
employ a convolutional neural network (CNN)
\citep{Fukushima:1980,LeCun:1989} --- a specialized kind of neural
  network for processing data that has a grid-like
structure. It is widely used in the pattern recognition and image
classification problems. In recent years CNNs have seen many applications in
physics~\citep{Carleo:2019ptp} and astronomy
\citep[see, for example,][]{Kim:2016knv, Petrillo:2017, Hezaveh:2017sht,
  Vernardos:2019}. In this section we will show that CNN is able to detect
glitches of gamma-ray pulsars with high accuracy. This gives opportunity
for glitches identification in an extensive searches dealing with large amount
of data, when it can't be done manually.

The architecture of the CNN used in this work is presented in
  Table~\ref{table1}. It was implemented in Python using
  Keras~\citep{Chollet} library with the Tensorflow backend. The network
  has two components: the feature extraction part and the classification part.
  The feature extraction part consist of convolution and polling layers
  which detect specific patterns for pulsars of both kinds. The fully
  connected layers constitute the classification part. It assign a
  probabilities that the input data corresponds to glitching
  (non-glitching) pulsar.

\begin{figure}
  \resizebox{\hsize}{!}{\includegraphics{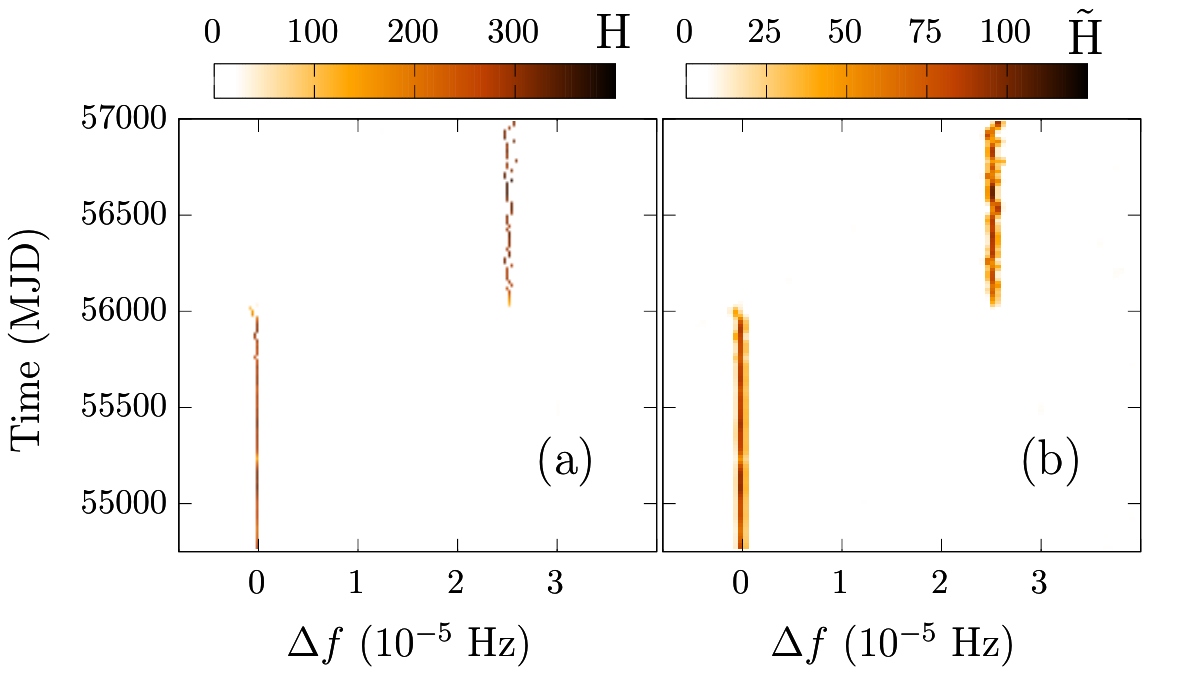}}
  \caption{(a) The weighted $H$-test data calculated as we discussed in
    Section~\ref{sec:method} for generated pulsar with glitch and (b) this
    data after convolution with Gaussian function~\eqref{Hs}.}
  \label{fig3}
\end{figure}

The weighted $H$-test dependence on frequency and time is used for glitches
recognition. It is calculated as we have discussed in Section~\ref{sec:method}.
Before being fed to the CNN the data is convolved with a Gaussian
function according to the formula
\begin{align}
\label{Hs}
\tilde{H}(f,t) = \int_{-\infty}^{\infty} d\tilde{f}\,H(\tilde{f},t) \cdot
\frac{1}{\sqrt{2\pi} \delta_f} \mathrm{e}^{ -\frac{(f-\tilde{f})^2}{2\delta_f^2}}\;.
\end{align}
Search for large glitches generally requires to execute more steps in the scan
over frequency. The latter considerably increases the size of the resulting
$H$-test array. Thinning in this case will result in loss of high
$H$-test values corresponding to some narrow frequency bandwidth. The
convolution~\eqref{Hs} smooths small scale details spreading the $H$-test
values over the scales $f \sim \delta_f$. This allows to reduce the size
of the array $\tilde{H}$ keeping an average information. The
convolution~\eqref{Hs} can be calculated at any frequencies within the
range covered by the original data. In what
follows we compute convolution at $131$ equidistant frequency values for
every time group of the photons. As a result we obtain the array of the size
$131 \times 131$, which is fed to the CNN (see Table~\ref{table1}).
The foregoing can be seen in Figure~\ref{fig3}, where the original
data (see Figure~\ref{fig3}a) and the results of convolution
(see Figure~\ref{fig3}b) are shown.

The CNN predicts whether the input data corresponds to pulsar with glitch
or not. The output layer with sigmoid activation function returns a number
between $0$ and $1$. If the output is less than $0.5$ we assign the input
sample to the pulsars without glitch, otherwise we assign it to the pulsars
with glitches.

\begin{table}
\caption{Architecture of the convolutional neural network}
\label{table1}
\centering
\begin{tabular}{c c c}
\hline\hline
  No   & Layer          & Size output \\
  \hline
  1  & Input          & 131$\times$131$\times$1 \\
  2  & Conv2D         & 129$\times$129$\times$16 \\
  3  & MaxPooling2D   & 64$\times$64$\times$32  \\
  4  & Conv2D         & 62$\times$62$\times$32 \\
  5  & MaxPooling2D   & 31$\times$31$\times$32  \\
  6  & Conv2D         & 29$\times$29$\times$32 \\
  7  & MaxPooling2D   & 14$\times$14$\times$32  \\
  8  & Conv2D         & 12$\times$12$\times$32 \\
  9  & MaxPooling2D   & 6$\times$6$\times$32  \\
  10 & Conv2D         & 4$\times$4$\times$32 \\
  11 & MaxPooling2D   & 2$\times$2$\times$32 \\
  12 & Flatten        & 128 \\
  13 & Dropout        & 128 \\
  14 & Dense          & 128 \\
  15 & Dense          & 1 \\
\hline
\end{tabular}
\end{table}

The network contains a large number of parameters, tuning which
during the training stage requires a large data set. The amount of training
data qualifies the ability of the network to identify glitches. Since there
are not so many gamma-ray pulsars with glitches known whose can be employed
to train the network, we generated them as follows. First of all, we generate
randomly the pulsar frequency and spin-down rate within the ranges
$1$ Hz $\leq f \leq 10$ Hz and
$-10^{-15}\;\text{Hz}\cdot\text{s}^{-1}\leq \dot{f} 
\leq -10^{-13}\;\text{Hz}\cdot\text{s}^{-1}$ correspondingly.
Secondly, we introduce the pulsar light curve as a gaussian peak over 
a constant background level corresponding to the off-pulse emission. The width
of a peak is generated randomly from $0.05$ to $0.45$ of the pulsar period 
$2 \pi/f$. The value of the constant background is generated  
from $0.1$ to $0.6$ of the peak height. For the pulsars with glitches we also 
generate the time after which the pulsar frequency and spin-down 
rate get increments in the ranges $10^{-8} \leq \Delta f /f 
\leq 10^{-5}$ and $10^{-4} \leq \Delta \dot{f} /\dot{f} \leq 10^{-3}$ 
correspondingly. Finally, we generate randomly $40000$ photons with unit
weights according to this light curve with the barycentric arrival times from
54682 MJD to 57084 MJD which corresponds to $6.5$ years of observations.

The data set of $15000$ pulsars with glitches and $13500$ without glitch
was generated. For each of the generated sources the weighed $H$-test data are
calculated as discussed in Section~\ref{sec:method}. The range of the scan
over frequency is taken according to the glitch amplitude $\Delta f$ 
and randomly for pulsars without glitches. Then the convolution~\eqref{Hs} is
calculated. The results of applying the method to an example of generated
pulsar with a glitch are illustrated in Figure~\ref{fig3}.

The data are splitted randomly into two subsets:
$90\%$ of the data is for the training, $10\%$~---~for the validation.
To increase the amount of training data we use the augmentation
  techniques. We randomly crop a region around pre-glitch frequency value 
  from the original array $H(f,t)$ and after convolution with Gaussian
  function~\eqref{Hs} obtain $\tilde{H}(f,t)$.  The cross-entropy was
used as the loss function assuming the target value of $1$ for all samples
of pulsars with glitches and $0$ otherwise. The network was trained during
500 epochs while the overfitting was reduced by including a dropout layer
and using $L2$ regularization of the weights in the convolutional layers. 

\begin{figure}
  \resizebox{\hsize}{!}{\includegraphics{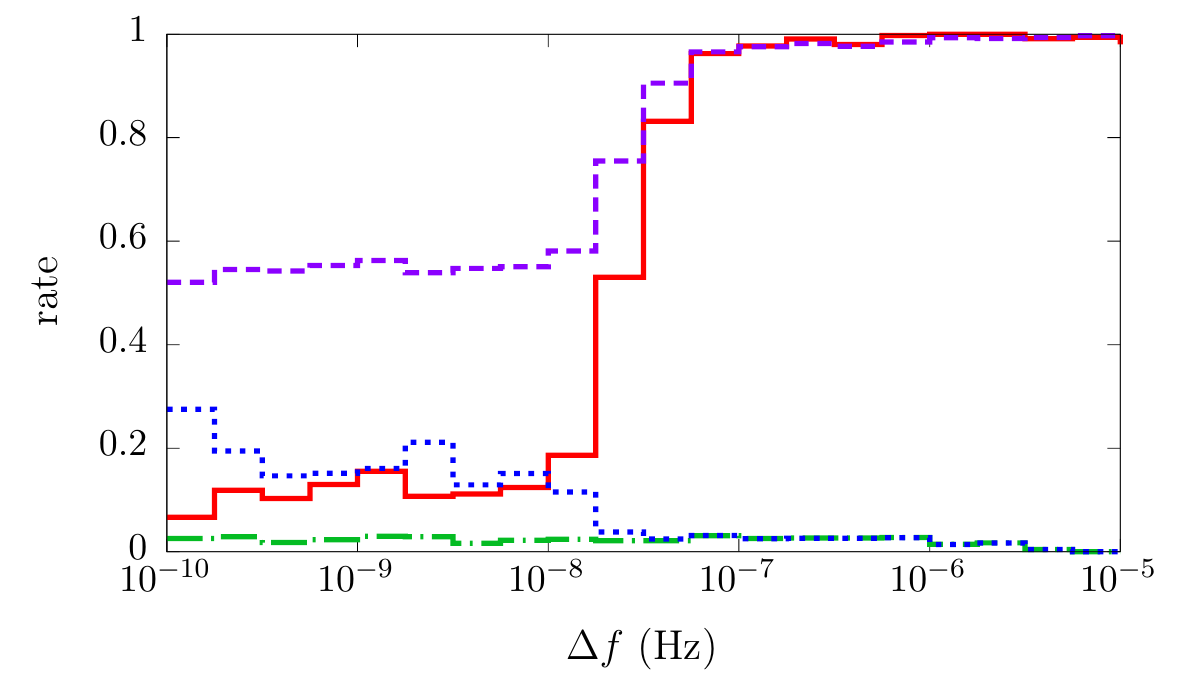}}
  \caption{The neural network efficiency of pulsar glitches detection:
    true positive rate (red solid line)~---~the proportion of correctly
    identified pulsars with glitch among all pulsars with glitch of the
    amplitude $\Delta f$, false positive rate (green dashed-dotted
    line)~---~the proportion of erroneously identified as pulsars with
    glitch among pulsars without glitches in the search for glitches of
    the amplitude $\Delta f$, accuracy (violet dashed line)~---~the
      proportion
    of correct predictions and false discovery rate (blue dotted line)~---~
    the proportion of erroneously identified as pulsars with glitch among
    all glitch identifications.
    \label{fig4}}
\end{figure}

\begin{figure}
  \resizebox{\hsize}{!}{\includegraphics{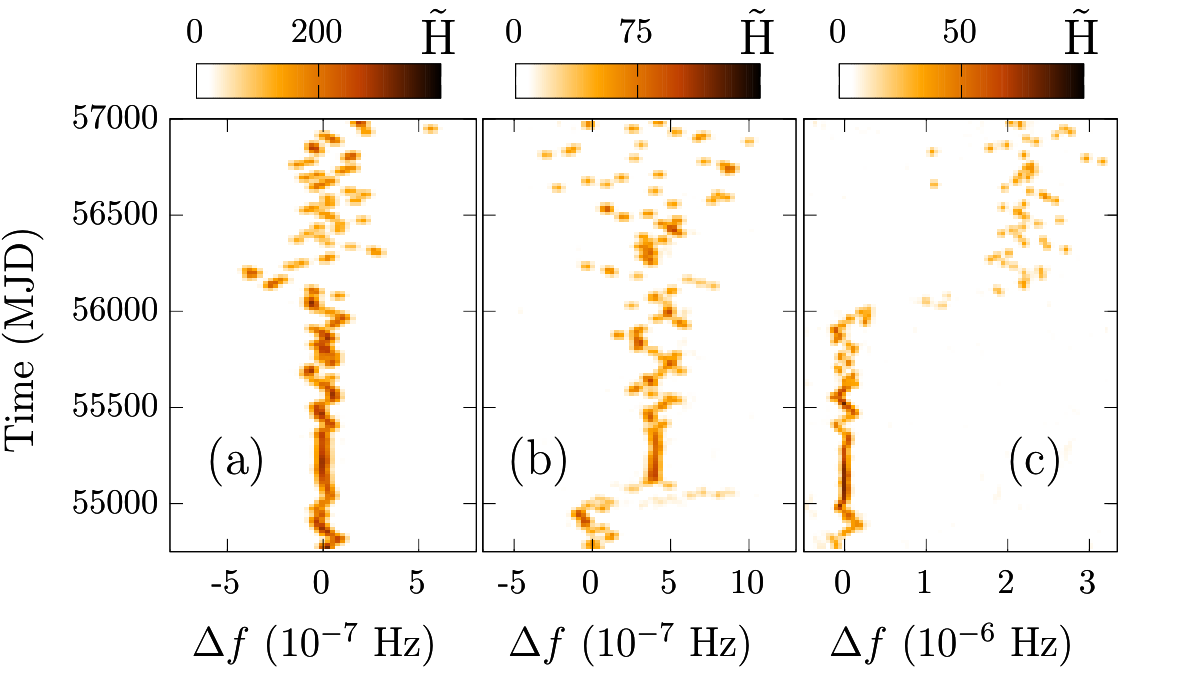}}
  \caption{The weighted $H$-test data calculated as we discussed in
    Section~\ref{sec:method} and convolved with Gaussian function~\eqref{Hs}
    for three generated pulsars with glitch:
    (a) at 56183 MJD with $\Delta f \simeq 4.6\times 10^{-8}$ Hz,
    (b) at 55045 MJD with $\Delta f \simeq 4.2\times 10^{-7}$ Hz and
    (c) at 56041 MJD with $\Delta f \simeq 2.2\times 10^{-6}$ Hz
    \label{fig5}
    }
\end{figure}

In order to test the accuracy of the CNN we generate additional $2500$ pulsars
without glitches and $2500$ pulsars with glitches of the amplitude
$10^{-10}$ Hz $\leq \Delta f \leq 10^{-5}$ Hz. The network performance on the
test data is presented in Figure~\ref{fig4}. As one can see from the figure,
the CNN demonstrates high accuracy in detection of large glitches.
About $98\%$ of pulsars with frequency change $\Delta f \gtrsim 10^{-7}$ Hz
were correctly identified. The false positive rate is about $2\%$ over the
whole frequency interval of the glitches search. The example of pulsars with
glitch correctly identified by the neural network is presented in
Figure~\ref{fig5}.

The fraction of correctly identified pulsars with glitch gradually
decreases with decreasing amplitude of the frequency shift reaching
approximately $10\%$ for $\Delta f \lesssim 10^{-8}$ Hz. We checked that an
extension of the training data set with pulsars with such small glitches does
not improve the accuracy. This means that the method has reached the threshold
of sensitivity, which corresponds to a resolution of the 170-day time
window $1/\Delta t \simeq 7 \times 10^{-8}$ Hz. One can increase the window
size, but this does not seem to improve considerably the sensitivity to such
small glitches.

The relative number of pulsars with glitch detected by the neural network below
the sensitivity threshold was expected to be at the same level as the false
positive rate. However, as one can see from Figure~\ref{fig4}, this fraction
is about $10\%$ which is much higher than $2\%$. The reason is in the
non-negligible change in the spin-down rate during glitch which was
generated in the range $10^{-4} \leq \Delta \dot{f} /\dot{f} \leq 10^{-3}$.
The weighted $H$-test calculated according to the
equations~\eqref{time},~\eqref{H} was maximized over
the spin-down rate leaving dependence on $t$ and $f$. However, some information
about change in $\dot{f}$ due to a glitch is likely to remain and is
“noticed” by the neural network. In order to test this hypothesis, we generate
two sets of 100 pulsars in each with the same glitch amplitude
$\Delta f = 10^{-10}$ Hz but with different changes of the spin-down rate
$\Delta \dot{f} = 10^{-20}\;\text{Hz}\cdot \text{s}^{-1}$ and
$\Delta \dot{f} = 5\times 10^{-17}\;\text{Hz}\cdot \text{s}^{-1}$ respectively.
The other parameters are fixed and the same in the each set. The result of
applying the neural network to these two sets confirmed the hypothesis:
in the set with $\Delta \dot{f} = 10^{-20}\;\text{Hz}\cdot \text{s}^{-1}$ two
pulsars were correctly identified while in another set the CNN
identified 11 samples as pulsars with glitches. 

\section{Discussion}

In this paper we have shown that the convolutional neural network applied to
the weighted $H$-test data can be used to detect glitches of the gamma-ray
pulsars in an automatic regime. The neural network demonstrates a very high
accuracy on the generated data and recognizes pulsars with glitches up to very
small amplitudes $\Delta f \sim 10^{-8} \text{Hz}$. It opens up a new
possibility to exploit this method in an extensive searches dealing with
large amount of data.

To verify that the network is able to recognize glitches in real data we 
apply it to some gamma-ray pulsars from the {\it Fermi}-LAT 3FGL catalog.
The neural network correctly identified pulsars with glitch known previously
including those presented in Figures~\ref{fig1},~\ref{fig2}. We postpone an
extensive searches of new glitches with the fine-tuning of coordinates for
the future works.

It is worth emphasizing that we have not yet applied the neural network to a
large volume of real data, which can turn out to be more complicated for
glitches recognition. In the latter case the following improvements of the
method are possible. First of all, the scan over source coordinates will be
able to recover phase-coherence what will increase the sensitivity of the
method. Second, some features of the real data which confuse the network can
be replicated in the generation of the training data set. It will allow the
network to ''learn'' these features and make less mistakes.

\begin{acknowledgements}
  We are indebted to O.E. Kalashev, G.I. Rubtsov and Y.V. Zhezher for numerous
  inspiring discussions. The work is supported by the Russian Science Foundation
  grant 17-72-20291. The numerical part of the work is performed at the cluster
  of the Theoretical Division of INR RAS.
\end{acknowledgements}


\begin{thebibliography}{}

\bibitem[Abdo at al. (2009)]{Abdo:2009}
Abdo, A.~A., Ackermann, M., Ajello, M., et al., 2009,
Sci, 325, 840
  
\bibitem[Abdo at al. (2013)]{TheFermi-LAT:2013ssa}
Abdo, A.~A., Ajello, M., Allafort, A., et al., 2013,
\apjs, 208, 17

\bibitem[Acero at al. (2015)]{3FGL}
Acero, F., Ackermann, M., Ajello, M., 2015, 
\apjs, 218, 2, 41

\bibitem[Atwood at al. (2009)]{Atwood}
  Atwood, W. B., Abdo, A. A., Ackermann, M., et al. 2009,
  \apj, 697, 1071

\bibitem[Ball\& Brunner(2010)]{BallBrunner}
  Ball, N.~M., \& Brunner, R.~J., 2010,
Int.J.Mod.Phys.D, 19, 07, 1049

\bibitem[Baron (2019)]{Baron}
Baron, Dalya, 2019,
\eprint arXiv:\href{https://arxiv.org/pdf/1904.07248}{1904.07248}

\bibitem[Bertsch et al. (1992)]{Bertsch}
  Bertsch, D.~L., Brazier, K.~T.~S., Fichtel, C.~E., et al., 1992,
 \nat, 357, 306

\bibitem[Carleo et al. (2019)]{Carleo:2019ptp}
  Carleo, G., Cirac, I., Cranmer, K., et al.,2019,
  Rev. Mod. Phys., 91, 4, 045002

  
\bibitem[Chollet et al. (2015)]{Chollet}
 Chollet, F., et al., 2015, 
\href{https://github.com/keras-team/keras}{Keras}


\bibitem[Clark et al. (2015)]{Clark:2015}
Clark, C.~J., Pletsch, H.~J., Wu, J., et al., 2015,
\apj, 809, 1, L2

\bibitem[Clark et al. (2017)]{Clark:2017}
Clark, C.~J., Wu, J., \& Pletsch, H.~J., 2017,
\apj, 834, 2, 106

\bibitem[Cordes et al. (1988)]{Cordes}
  Cordes J.~M., Downs G. S. \& Krause-Polstorff J., 1988,
  \apj, 330, 847
  
\bibitem[de Jager \& Busching (2010)]{deJager:2010ci}
de Jager, O.~C., \& Busching, I., 2010,
\aap, 517, L9

\bibitem[de Jager et al. (1989)]{Jager:1989}
  de Jager, O.~C., Raubenheimer, B.~C., \& Swanepoel, J.~W.~H., 1989,
\aap, 221, 180

\bibitem[Eatough et al.(2010)]{Eatough}
Eatough, R.~P., Molkenthin, N.,\& Kramer, M., 2010,
\mnras, 407, 4, 2443
\bibitem[Espinoza et al.(2011)]{Espinoza}
  Espinoza, C.~M., Lyne, A.~G., Stappers, B.~W.,\& Kramer M., 2011,
  \mnras, 414, 2, 1679

\bibitem[Espinoza et al. (2014)]{Espinoza:2014}
  Espinoza, C.~M., Antonopoulou, D., Stappers, B.~W., Watts, A., \& Lyne A.~G., 2014,
 \mnras, 440, 3, 2755


\bibitem[Fukushima (1980)]{Fukushima:1980}
 Fukushima K., 1980,
 Biological cybernetics, 36, 193

 
\bibitem[Halpern \& Holt (1992)]{Halpern}
  Halpern, J.~P., \& Holt, S.~S., 1992,
  \nat, 357, 222

\bibitem[Haskell \& Melatos, (2015)] {Haskell:2015jra}
Haskell, B., \& Melatos, A., 2015,
Int. J. Mod. Phys. D, 24, 3, 1530008

\bibitem[Hezaveh et al. (2017)]{Hezaveh:2017sht}
Y.~D.~Hezaveh, L.~Perreault Levasseur, \& P.~J.~Marshall, 2017
\nat, 548 (2017), 555-557

\bibitem[Kim \& Brunner, (2017)]{Kim:2016knv}
Kim, E.~J., \& Brunner, R.~J, 2017,
\mnras, 464, 4, 4463-4475

\bibitem[LeCun et al. (1989)]{LeCun:1989}
Le Cun, Y., Guyon, I., Jackel, L.D., et al., 1989,
Communications Magazine, 27(11), 41-46

\bibitem[Li et al. (2016)]{Li:2016uxb}
Li, J., Torres, D.~F, de Ona Wilhelmi, E., Rea, N., \& Martin, J., 2016,
\apj, 831, 1, 19

\bibitem[Lyne et al. (2000)]{Lyne:2000}
  Lyne A. G., Shemar S. L., Graham-Smith F., 2000,
  \mnras, 315, 534

\bibitem[Lyne et al. (2015)]{Lyne}
  Lyne A. G., Jordan C. A., Graham-Smith, F., et al., 2015,
  \mnras, 446, 857

\bibitem[McKee (2016)]{McKee}
 McKee, J.~W., Janssen, G.~H., \& Stappers, B.~W., et al., 2016,
 \mnras, 461, 3, 2809

\bibitem[McKenna \& Lyne, (1990)] {McKenna}
  McKenna J., Lyne A. G., 1990,
  \nat, 343, 349

\bibitem[Packard (1972)]{Packard:1972}
Packard, R. E., 1972,
Physical Review Letters, 28, 1080

\bibitem[Packard et al. (2018)]{Packard:2018}
  Palfreyman J., Dickey J. M., Hotan A., Ellingsen S., \& van Straten W., 2018,
  \nat, 556, 219

\bibitem[Petrillo et al. (2017)]{Petrillo:2017}  
  Petrillo, C. E., Tortora, C., Chatterjee, S., et al, 2017,
\mnras, 472, 1, 1129–1150  
  
\bibitem[Pletsch et al. (2012)]{Pletsch:2012}
Pletsch, H.~J., Guillemot, L., Allen, B., et al., 2012,
\apjl, 755, 1, L20
 
\bibitem[Pletsch et al. (2013)]{Pletsch:2013}
Pletsch, H.~J., Guillemot, L., Allen, B., et al., 2013,
\apj, 779, L11

\bibitem[Radhakrishnan \& Manchester (1969)]{Manchester}
  Radhakrishnan, V., \& Manchester, R.~N., 1969,
\nat, 222, 228

\bibitem[Reichley \& Downs (1969)]{Downs}
  Reichley, P.~E., \& Downs, G.S., 1969,
\nat, 222, 229  

\bibitem[Ruderman (1969)]{Ruderman}
Ruderman, M., 1969, 
\nat, 223, 597

\bibitem[Saz Parkinson et al. (2010)]{Saz:2010}
Saz Parkinson, P. M., Dormody, M., Ziegler, M., et al., 2010, 
\apj, 725, 571

\bibitem[Shannon et al. (2016)]{Shannon}
  Shannon R. M., Lentati L. T., \& Kerr, M., et al., 2016,
  \mnras, 459, 3104

\bibitem[Sokolova \& Rubtsov (2016)]{Sokolova:2016btd}
Sokolova, E., \& Rubtsov, G., 2016,
\apj, 833, 2, 271

\bibitem[Vernardos \& Tsagkatakis (2019)]{Vernardos:2019}
Vernardos, G., \& Tsagkatakis, G., 2019,
\mnras, 486, 2, 1944–1952

\bibitem[Vivekanand (2017)]{Vivekanand}
 Vivekanand, M., 2017, 
 \eprint arXiv:\href{https://arxiv.org/pdf/1710.05293}{1710.05293}
 
\bibitem[Yu et al. (2013)]{Yu:2013}
  Yu, M., Manchester R.~N., Hobbs, G., et al., 2013,
  \mnras, 429, 1, 688–724

\end{thebibliography}
\end{document}